# Epistasis not needed to explain low dN/dS

David M. McCandlish[1], Etienne Rajon[1], Premal Shah[1], Yang Ding[1], and Joshua B. Plotkin[1,*]

[1]Department of Biology, University of Pennsylvania, Philadelphia, PA 19104

[*]Corresponding author: jplotkin@sas.upenn.edu

An important question in molecular evolution is whether an amino acid that occurs at a given position makes an independent contribution to fitness, or whether its effect depends on the state of other loci in the organism's genome, a phenomenon known as epistasis[1-5]. In a recent letter to Nature, Breen et al.[6] argued that epistasis must be "pervasive throughout protein evolution" because the observed ratio between the per-site rates of non-synonymous and synonymous substitutions (dN/dS)[7] is much lower than would be expected in the absence of epistasis. However, when calculating the expected dN/dS ratio in the absence of epistasis, Breen et al. assumed that all amino acids observed in a protein alignment at any particular position have equal fitness. Here, we relax this unrealistic assumption and show that any dN/dS value can in principle be achieved at a site, without epistasis. Furthermore, for all nuclear and chloroplast genes in the Breen et al. dataset, we show that the observed dN/dS values and the observed patterns of amino acid diversity at each site are jointly consistent with a non-epistatic model of protein evolution.

For a variety of proteins under long-term purifying selection, Breen et al. constructed alignments and recorded the amino acids observed at each position; these observed amino acids were deemed to be "acceptable" with respect to natural selection. Breen et al. then assumed that substitutions occur at neutral rates among these acceptable amino acids, and used these rates to calculate, for each protein, an expected value for dN/dS in the absence of epistasis. Because their empirical observations of dN/dS were much lower than these expected values, Breen et al. concluded that epistasis must be extremely prevalent.

The flaw in this reasoning is that Breen et al. considered only a single class of fitness assignments, such that all amino acids observed at a site were assumed equally fit. However, a more realistic alternative is simply that some amino acids observed at a site are more fit than others[8,9].

As a proof of principle, we considered a non-epistatic model in which, among the acceptable amino acids at a given site, one of these is preferable to the rest. We performed the following experiment: in a hypothetical protein of length 300 aa, for each position we randomly designated 8 amino acids as acceptable (the average number of acceptable amino acids reported by Breen et al.) but gave one of these a selective advantage over the rest. We then calculated the equilibrium dN/dS[10] for this protein as a function of the selective advantage of the preferred amino acid at each site, 2Ns (Figure 1). While dN/dS is high for the case 2Ns=0, corresponding to the Breen et al. assumption, dN/dS is much lower for larger 2Ns. Thus, a large range of dN/dS values are consistent with non-epistatic models of protein evolution.

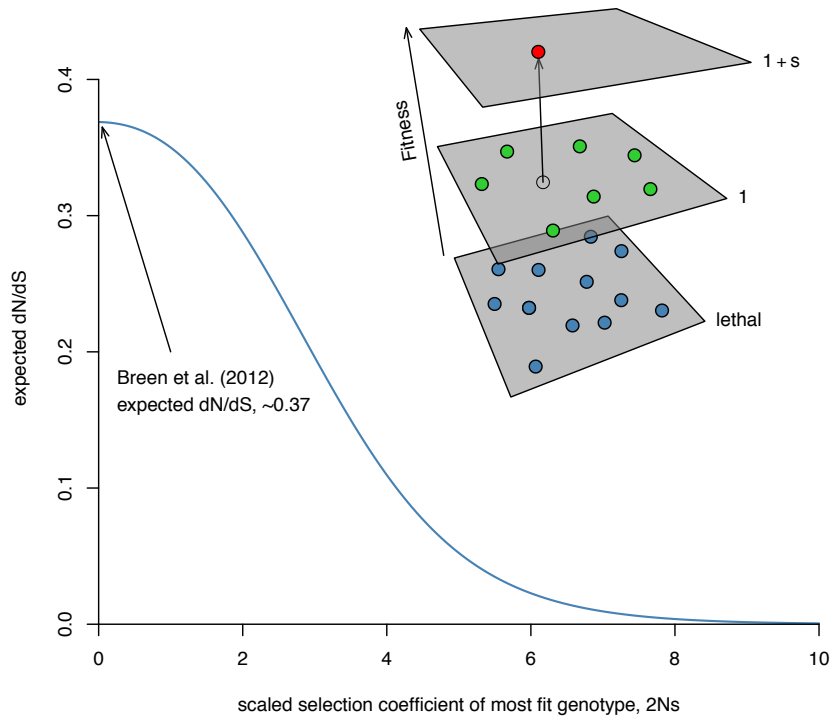

**Figure 1: dN/dS and the strength of selection.** dN/dS as a function of 2Ns for a hypothetical protein of length 300, in which 8 acceptable amino acids are chosen at random for each position and one of these amino acids at random is assigned a selective advantage of size 2Ns. The remaining 12 amino acids are lethal. The Breen *et al.* expectation for dN/dS in the absence of epistasis corresponds to 2Ns=0.

Although site-independent fitness assignments can in principle produce very low dN/dS values, are the Breen *et al.* data in fact consistent with a non-epistatic model? To answer this question, for each protein in the Breen *et al.* dataset, we assigned amino acids fitnesses at each site so that their equilibrium frequencies precisely matched the frequencies observed by Breen *et al.* for that site[11]. For each protein, we then simulated the divergence between pairs of sequences and computed dN/dS for each pair. For the mitochondrial genes in the Breen *et al.* dataset, the average simulated dN/dS values, while on the whole substantially lower than the Breen *et al.* expectations, were still greater than the empirically observed values. However, for all nuclear and chloroplast genes in the Breen *et al.* dataset, the average dN/dS values under our non-epistatic model are comparable to or even lower than the empirical dN/dS values reported by Breen *et al.* Thus, the low dN/dS in these genes need not be attributed to epistasis (Table 1), but rather can be explained by the more parsimonious assumption that some amino acids are more fit than others, at each site.

| Gene | Breen et al. Expected dN/dS | Our Average Simulated dN/dS | Breen et al. Empirical dN/dS |
|---|---|---|---|
| **Mitochondrial** | | | |
| ATP6 | 0.44 | 0.215 | 0.056 |
| ATP8 | 0.56 | 0.624 | 0.224 |
| COX1 | 0.28 | 0.078 | 0.015 |
| COX2 | 0.43 | 0.140 | 0.025 |
| COX3 | 0.32 | 0.144 | 0.036 |
| CYTB | 0.51 | 0.117 | 0.039 |
| ND1 | 0.39 | 0.208 | 0.040 |
| ND2 | 0.51 | 0.262 | 0.067 |
| ND3 | 0.49 | 0.242 | 0.069 |
| ND4 | 0.42 | 0.239 | 0.045 |
| ND4L | 0.49 | 0.369 | 0.076 |
| ND5 | 0.32 | 0.211 | 0.057 |
| ND6 | 0.42 | 0.397 | 0.073 |
| **Nuclear** | | | |
| EEF1A1 | 0.11 | 0.031 | 0.020 |
| H3.2 | 0.14 | 0.014 | 0.037 |
| **Chloroplast** | | | |
| rbcL | 0.40 | 0.024 | 0.072 |

**Table 1: Observed and expected dN/dS.** Comparison of expected dN/dS values with the empirical values for each gene in the Breen *et al.* dataset. The Breen *et al.* expected dN/dS is based on the assumption that all amino acids observed at a given site are neutral relative to each other. Our average simulated dN/dS is based on the assumption that the amino acids observed at a given site have different fitnesses; these fitnesses are chosen so that the equilibrium amino acid frequencies at each site match the empirical frequencies at that site in the Breen *et al.* dataset.

It is important to note that the effects of natural selection and phylogeny are confounded in the observed amino acid frequencies. For small datasets, and with a known phylogeny, methods exist to distinguish these effects[12,13]. However, in the absence of a phylogeny, our fitness estimates are maximum-likelihood, and in equilibrium they are guaranteed to reproduce both the site-specific amino acid frequencies and the mean pairwise divergences between the empirical amino acid sequences. Our method makes the standard assumption (e.g. ref. 14, as used by Breen *et al.*) that molecular evolution can be modeled as an equilibrium Markov chain running on the branches of a phylogeny. Relaxing this assumption would make it even more difficult to reject the non-epistatic null hypothesis, because it would then be necessary to rule out the even larger class of non-equilibrium site-independent models before concluding that epistasis is present.

In summary, the data of Breen *et al.* do not support the conclusion that epistasis is the primary factor in molecular evolution. Breen *et al.* provide no direct evidence of epistasis whatsoever, but rather their conclusions were based on the rejection of an overly simplistic model of site-independent evolution. At best, the data support the conclusion that epistasis may be an important factor in the evolution of mitochondrial genes, but to a lesser extent than originally claimed.

Methods:

We assume that each codon evolves according to an independent Markov chain whose rate matrix is determined by the scaled selection coefficient assigned to each amino acid[15]. The equilibrium frequency of each amino acid is then proportional to $u_i e^{2Ns(i)}$ (ref. 15) where $u_i$ is the number of codons that code for amino acid $i$ and $2Ns(i)$ is its scaled selection coefficient. Simulations were conducted by running an independent Markov chain for each codon represented in an at least half the sequences of the Breen *et al.* alignment, until the divergence between each pair of sequences was dS=0.25, which is within the range of dS=0.05 to 0.5 used by Breen *et al.* dN/dS was then estimated using PAML[14], again following the procedure described by Breen *et al.* The ancestral sequence at each site was drawn from the equilibrium distribution. All computer codes can be made available upon request.